\documentstyle[preprint,prb,aps]{revtex}
\begin{document}
\draft
\title{$XY$ universality of the stacked triangular Ising antiferromagnet}
\author{M.L. Plumer}
\address{Centre de Recherche en Physique du Solide et D\'epartement de
Physique}
\address{Universit\'e de Sherbrooke, Sherbrooke, Qu\'ebec, Canada J1K 2R1}
\author{A. Mailhot}
\address{INRS-EAU, 2800 rue Einstein, C.P. 7500,
              Ste.-Foy, Qu\'ebec, Canada G1V 4C7}
\date{July 1994}
\maketitle
\begin{abstract}
Histogram analysis of Monte-Carlo simulation results
of the Ising antiferromagnet on a stacked triangular lattice are reported.
Finite-size
scaling estimates for critical exponents suggest $XY$ universality.
This result supports earlier predictions based on symmetry arguments
and is in contrast with a recent suggestion of a new universality
class by Bunker {\it et al.} [Phys. Rev. B {\bf 48}, 15861 (1993)].
\end{abstract}
%============================================================================
% BODY OF PAPER

Controversy regarding the critical behavior of frustrated spin systems
continues.\cite{diep}  The difficulty in resolving such issues is due
partly to the disparity in results, and in their interpretation, among the
wide variety of theoretical and numerical approaches available.  Even
Monte-Carlo (MC) simulations performed by two different groups can give
results which lead to two different conclusions.  Sets of critical exponents
associated with different universality classes may differ by only a
relatively small amount.  In the case of the $XY$ antiferromagnet on a
stacked triangular lattice (STL), recent histogram MC results have
produced exponent estimates,\cite{plum1} not too different from those
found earlier by Kawamura based on conventional MC analysis.\cite{kawa}  In the
former case, however, the results are suggestive of mean-field tricritical
behavior in contrast with the proposal by Kawamura of a new universality
class.  Although MC simulation results by a number of groups in the Heisenberg
case concur with those found by Kawamura (see, e.g.,
Ref.\onlinecite{heis}), thus supporting the scenario of a new universality
class for that model system, these data do not agree with the proposal by
Azaria {\it et al.}\cite{aza}
of either $O(4)$ or tricritical universality (if the transition is continuous).
In the present work, finite-size scaling of extensive histogram MC data
in the Ising case is shown to yield critical exponents
which strongly suggest $XY$ universality.  This
result supports the symmetry arguments made some ten years ago by
Berker {\it et al.}\cite{berker}
and is in contrast with the recent suggestion of a new universality
class for this system.\cite{bunk}

The finite-size scaling of histogram MC data can yield highly accurate
critical exponents  for unfrustrated models.\cite{pec2,ferr}
For a system of size $L \times L \times L$, exponent ratios determine
the scaling with $L$ of both the extrema of thermodynamic functions
(except the order parameter) near
the critical temperature, $T_N$, and also these functions evaluated {\it at}
$T_N$.  The latter procedure appears to be both more accurate (especially
in the case of the susceptibility) and less
computer-time expensive with use of the cumulant crossing
method to obtain reliable estimates of critical
temperatures.\cite{pec2}  With the former procedure,
histograms must be made at temperatures close to the where the extrema
occur, which can vary significantly with $L$.  The multiple-histogram
technique is useful in this case, but with more computational expense.

Frustrated systems are expected to exhibit more fluctuations due
to the proximity of nearly degenerate states.\cite{diep}  Although
much progress has been made in recent years to overcome critical
slowing down, these techniques appear largely ineffective on
frustrated systems.  Special attention must then be given to
ensure that MC runs are sufficiently long and that $L$ is sufficiently
large for the system to exhibit its true critical behavior.\cite{plum2}

The present work serves to complement and extend our previous results
on the STL Ising antiferromagnet which focused on the effects of
next-nearest-neighbor basal-plane coupling.\cite{plum3}  With these limited
data (i.e., only $1 \times 10^6$ MCS (Monte-Carlo steps) were used for
averaging at each $L$=12-30), the
exponents were estimated by the scaling of extrema of thermodynamic
functions also for the near-neighbor model.  Given the rather large
uncertainties in these results, only {\it consistency} with $XY$ universality
could be claimed.

In contrast, Bunker {\it et al.}\cite{bunk} employed the multiple histogram
technique to estimate exponents from the scaling of extrema and, although
their results are not too far from those of $XY$ universality, the
differences were significant enough to lead to the suggestion that the STL
Ising
antiferromagnet belongs to a new universality class.
Their histograms were made at $L$=6-30, with averaging made using
approximately $1 \times 10^6$ MCS for the smaller lattices and only
$2.2 \times 10^5$ MCS at $L$=30.

In an effort to further test the hypothesis of a new universality class,
we report here the results of extensive MC simulations performed on the STL
Ising model with antiferromagnetic near-neighbor exchange coupling
in the basal plane, $J_\bot = 1$, and ferromagnetic coupling along
the $c$-axis, $J_\| = -1$.
The Metropolis MC algorithm was used to generate histograms
on lattices with $L$=12-33 using runs with from
$5 \times 10^5$ MCS for the smaller lattices to $1.2 \times 10^6$ MCS for the
larger lattices, after discarding the initial
$1 \times 10^5$ - $2 \times 10^5$
MCS for thermalization.  Averaging was then made using from 6 runs
for the smaller $L$ to 15 runs at $L$=30.  Averaging was made over
13 runs at $L$=33. For the largest lattice, this gives a reasonable
$15.6 \times 10^6$ MCS for averaging.
Errors were estimated (approximately) by taking the standard
deviation of the many runs for each $L$.
All histograms were generated at our previous estimate
of the critical temperature, $T_N \simeq 2.93$.\cite{plum3}

Results of applying the cumulant-crossing method\cite{pec2} to estimate the
critical temperature
are presented in Fig. 1.  The points represent the temperatures at which
the order-parameter cumulant $U_m(T)$ at $L'$ crosses the cumulant at
$L=12$ or $L=15$.
There is considerable scatter in the data and care must be taken to
use only results with $L$ sufficiently large to be in the
asymptotic region where a linear extrapolation is justified \cite{pec2},
i.e., for $Ln^{-1}(L'/L)$
\hbox{$<$\kern -0.8em\lower 0.8ex\hbox{$\sim$}} 2.2.
{}From these results, the critical temperature is estimated to be
$T_N=2.9298(10)$.  This value compares well with our previous estimate
but differs significantly from that of Bunker {\it et al.}, $T_N=2.920(5)$.

Finite-size scaling results at $T_N$ for the specific heat $C$, spin order
parameter
$M$, susceptibility (as defined in Kawamura's work \cite{kawa,pec2})
$\chi$, and the first logarithmic derivative of the order parameter
$V_1 = \partial [Ln(M)]/ \partial K$ (where $K=T^{-1}$)
\cite{ferr} are shown in Figs. 2-5.
Except in the case of the specific heat, exponent ratios were estimated
by assuming a scaling dependence $F=aL^x$ for a function of interest.
Errors were too large to extract a reliable estimate for $\alpha / \nu$
and Fig. 2 shows the scaling results with an
assumption of $XY$ universality.  A good straight-line fit is obtained
using data from the five largest lattices.  Estimates for the other exponent
ratios were made using either all the data, with the $L$=12 data excluded,
with $L$=12-15 data excluded, etc.
Only the estimates for $1 / \nu$ displayed a significant variation when data
for
the smaller lattices were not included in the fitting procedure.
These were made using $V_1$,
with the results 1.531(6), 1.526(5), 1.521(5), and 1.515(5), respectively
(quoted uncertainties represent the robustness of the fitting procedure
and does not include the effects of the error bars).
Corresponding results using the second logarithmic derivative\cite{pec2} $V_2$
are 1.536(6), 1.532(5), 1.527(5), and 1.521(4).
The sets of values represent fits from 8 (first numbers) to only 5
(last numbers) data points.   Although caution
must therefore be used in weighing the significance of these latter estimates,
the sets of values suggest that $1 / \nu \simeq 1.51(2)$ is a reasonable
extrapolation for the thermodynamic limit.

Scaling of estimated maxima of the function $V_1$ was also performed.
The result of using our data for $L$=12-30 is $1/\nu \simeq 1.46$, exactly the
value found by Bunker {\it et al.}.  With only results $L$=21-30
included, the estimate $1/\nu \simeq 1.44$ was obtained.  Since simulations
were all performed at only one temperature (T=2.93), the uncertainty associated
with these scaling results is large.  For example, the maxima in $V_1$ at
$L$=30 was found at $T=2.952$, quite far from the simulation temperature.  In
principle, larger values for the maxima in $V_1$ would be obtained if
the simulations were performed at temperatures closer to where they
occur.  This would lead to an increase in the value of $1/\nu$.  Similar
conclusions may be relevant in the case of the results at $L$=30
of Bunker {\it et al.}

The final values quoted for $\beta / \nu$ and $\gamma / \nu$ correspond
to fits made with the two smallest lattices excluded.
In order to estimate errors
due to the uncertainty in $T_N$, identical scaling was also performed
at $T=2.9288$ and $T=2.9308$.  Uncertainties arising from the error bars
indicated on the figures were always found to be smaller.
The resulting exponents (using  $\nu \simeq 1/1.51 = 0.662$) and error
estimates
are presented in Table I, along with a variety of
exponent estimates for the standard
$XY$ model, as well as those of Bunker {\it et al.}.  Our results are
well within the range of values found for the $XY$ model.

In conclusion, these results clearly indicate that the Ising
antiferromagnet on a
stacked triangular lattice exhibits a phase transition which belongs to
the standard $XY$ universality class.  This is in agreement with the
symmetry arguments of Berker {\it et al.}\cite{berker} and not with the
suggestion of a new universality class recently made by
Bunker {\it et al.}\cite{bunk}.  We believe that the critical exponents
estimated from the MC simulations reported in this latter work were made
using lattices that were too small and with insufficient statistics at
the larger lattice sizes.  The scaling of extrema of thermodynamic functions,
rather than at the critical point, may also reduce the accuracy of
exponent estimates.  These conclusions serve as a general reminder that
special care must be taken in the numerical simulation of frustrated spin
systems.

\acknowledgements
This work was supported by NSERC of Canada and FCAR du Qu\'ebec.
%==============================================================================
%

%==========================================================================
%TABLE

\vfill\eject
\begin{table}
\caption{ Comparison of estimates for critical exponents of the unfrustrated
$XY$ model and those for the stacked triangular (STL) Ising antiferromagnet.}
\vskip0.2cm
\begin{tabular}{ccccc}
                                & $\alpha$ & $\beta$ & $\gamma$ & $\nu$ \\
\tableline
XY: RG ($4-\epsilon$.)$^a$        &-0.013(15)&0.349(4) & 1.315(7) & 0.671(5) \\
XY: HT Series.$^b$                &-0.01(2)  &0.345(10)& 1.323(15)& 0.670(7) \\
XY: HT Series.$^c$                &-0.01(3)  &0.348(15)& 1.315(9) & 0.67(1)  \\
XY: Histogram MC.$^d$             &-0.010(6) &0.347(3) & 1.316(5) & 0.670(2) \\
XY: Histogram MC.$^e$             &+0.014(20)&0.331(10)& 1.324(1) & 0.662(7) \\
STL Ising: This work.$^f$         &+0.012(30)&0.341(4) & 1.31(3)  & 0.662(9) \\
STL Ising: Bunker {\it et al.}$^g$&-0.05(3)  &0.311(4) & 1.43(3)  & 0.685(3) \\
\end{tabular}
\tablenotetext[1]{Renormalization-group from Ref.\onlinecite{leg}
and scaling relations.}
\tablenotetext[2]{High-temperature series from Ref.\onlinecite{fer}
and scaling relations.}
\tablenotetext[3]{High-temperature series from Ref.\onlinecite{but}
and scaling relations.}
\tablenotetext[4]{From Ref.\onlinecite{janke} and scaling relations.}
\tablenotetext[5]{From Ref.\onlinecite{gott} and scaling relations.}
\tablenotetext[6]{With $\alpha$ determined from scaling relations.}
\tablenotetext[7]{From Ref.\onlinecite{bunk} with $\alpha$ determined
                             from scaling relations.}
\end{table}

%=============================================================================
%FIGURES

\begin{figure}
\caption{Results of applying the cumulant-crossing method (see text)
to estimate the critical temperature, where $b=L'/L$.}
\label{fig1}
\end{figure}

\begin{figure}
\caption{Finite-size scaling of the specific heat data.
Data at L=12 and L=15 is excluded from the fit.  Error bars are estimated
from the standard deviation found in the MC runs.}
\label{fig2}
\end{figure}

\begin{figure}
\caption{Finite-size scaling of the order parameter.
Data at L=12 and 15 are excluded from the fit.  Error bars are estimated
from the standard deviation found in the MC runs.}
\label{fig3}
\end{figure}

\begin{figure}
\caption{Finite-size scaling of the susceptibility $\chi$, as in Fig. 3.}
\label{fig4}
\end{figure}

\begin{figure}
\caption{Finite-size scaling of the logarithmic derivative of the order
parameter $V_1$ (see text), as in Fig. 3.}
\label{fig5}
\end{figure}

%==============================================================================
\end{document}